\documentstyle[12pt,epsf]{article}
\newlength{\abstractwidth}
\renewcommand{\thefootnote}{\fnsymbol{footnote}}
\renewcommand{\thanks}[1]{\footnote{#1}} 
\newcommand{\starttext}{
\setcounter{footnote}{0}
\renewcommand{\thefootnote}{\arabic{footnote}}}
\renewcommand{\theequation}{\thesection.\arabic{equation}}
\newcommand{\be}{\begin{equation}}
\newcommand{\bea}{\begin{eqnarray}}
\newcommand{\eea}{\end{eqnarray}}
\newcommand{\beq}{\begin{equation}}
\newcommand{\ee}{\end{equation}}
\newcommand{\eeq}{\end{equation}}

\newcommand{\<}{\langle}

\renewcommand{\b}{\beta}

\renewcommand{\>}{\rangle}
\def\ba{\begin{eqnarray}}
\def\ea{\end{eqnarray}}

\def\C{Complementarity}

\def\hp{Holographic Principle}
\def\14{{1\over4}}
\def\12{{1 \over 2}}

\def\h3{h^{3\over 2}}

\def\qft{quantum field theory}
\def\>{\rangle}
\def\<{\langle}
\def\sc {Schwarzschild}
\def\ls{\sqrt{\alpha'}}
\def\des{de Sitter Space}
\def\f{\Phi}

\begin{document}
\renewcommand{\theequation}{\thesection.\arabic{equation}}
\begin{titlepage}
\bigskip
\rightline{SU-ITP 02-11} \rightline{hep-th/0204027}

\bigskip\bigskip\bigskip\bigskip

\centerline{\Large \bf {Twenty  Years of Debate    }}
\centerline{\Large \bf {with Stephen }}

\bigskip\bigskip
\bigskip\bigskip

\centerline{\it L. Susskind  }
\medskip
\centerline{Department of Physics} \centerline{Stanford
University} \centerline{Stanford, CA 94305-4060}
\medskip
\medskip

\bigskip\bigskip
\begin{abstract}
This is  my contribution to Stephen Hawking's 60th birthday party.
Happy Birthday Stephen!
\medskip
\noindent
\end{abstract}

\end{titlepage}
\starttext \baselineskip=18pt \setcounter{footnote}{0}

\setcounter{equation}{0}
\section{Crisis and Paradigm Shift }

Stephen, as we all know, is by far the most stubborn and
infuriating person in the universe. My own scientific relation
with him I think can be called adversarial. We have disagreed
profoundly about deep issues concerning black holes, information
and all that kind of thing. At times he has caused me to pull my
hair out in frustration -- and you can plainly see the result. I
can assure you that  when we began to argue more than two decades
ago, I had a full head of hair.

I can also say that of all the physicists I have known he has had
the strongest influence on me and on my thinking. Just about
everything I have thought about since 1980 has in one way or
another been a response to his profoundly insightful question
about the fate of information that falls into a black hole
\cite{predict}. While I firmly believe his answer was wrong, the
question and his insistence on a convincing answer has forced us
to rethink the foundations of physics. The result is a wholly new
paradigm that is now taking shape. I am deeply honored to be here
to celebrate Stephen's monumental contributions and especially his
magnificent stubbornness.

The new paradigm whose broad outlines are already clear involves
four closely related ideas which I will call Horizon \C \ (also
known as Black Hole \C \ ) \cite{stretch,whiting} , the \hp \
\cite{hooft,holo}, the Ultraviolet/Infrared  connection
\cite{edlen} and the counting of black hole microstates
\cite{spec,sen,vafa,curt,gary}. Each has had strong support from
the mathematics of string theory. We have also learned that
certain lessons derived from \qft \ in a fixed background can lead
to totally wrong conclusions. For example \qft \ gives rise to an
ultraviolet divergent result for the entropy in the vicinity of a
horizon \cite{thooft,marks} and  a volume's worth of degrees of
freedom in a region of space. Another misleading result of \qft \
is that increasingly large energy and momentum scales are
equivalent to progressively smaller distance scales. Finally \qft
\ in a black hole background inevitably leads to the loss of
quantum coherence from the vantage point of a distant observer
\cite{predict}.

\setcounter{equation}{0}
\section{Stephen's Argument for Coherence  Loss }

Let's begin with the issue of coherence loss. Begin by drawing the
Penrose diagram for black hole formation. Let us think of it as a
background on which we can study conventional \qft . Now let's add
some additional particles which for simplicity are assumed
massless. Incoming particles enter the geometry on past light-like
infinity, $\cal{I}^-$. The initial Hilbert space of such particles
is labeled $\cal{H}_{-}$. The particles interact through Feynman
diagrams and some go out to $\cal{I}^+$ where they are seen by a
distant observer. But some particles end up at the space-like
singularity. A conventional local description would have the final
states live in a Hilbert space which is a product ${\cal{H}_+}
\bigotimes {{\cal{H}} _{sing}} $.

Applying ordinary rules to this system we expect the final state
to be given in terms of the initial state by an $S$ matrix, \be |f
\rangle = S |i \rangle \label{smatrix} \ee but to an observer
outside the black hole the final state consists only of the
particles on $\cal{I}^+$. It is therefore described by a density
matrix \be \rho=Tr_{sing}|f \rangle \langle f |. \label{dmatrix}
\ee Since black holes eventually evaporate we are left with a
mixed state and a loss of quantum coherence. I have simplified
some of the important parts of Stephen's  argument but in the
original paper he makes a very convincing case that \qft \ in a
black hole environment leads to a loss of coherence after the
evaporation has occurred. The problem with this is that loss of
quantum coherence violates the fundamental tenets of quantum
mechanics. That of course was Stephen's point.

Historically, for a long time high energy physicists, who had
other fish to fry, ignored  the problem. By and large, with one or
two exceptions \cite{dpage}, the relativity community accepted the
loss  of coherence without question. However,  't Hooft and I were
both deeply disturbed by Stephen's conclusion. In my case I felt
that loss of coherence, if it were to occur,  could not
quarantined     or isolated to
 phenomena which involved massive black holes. It would infect the rest
of physics
and cause dramatic disasters in ordinary situations \cite{bps}.
 't Hooft also expressed
similar concerns \cite{tHooftone}.

There was of course an alternative that I have not mentioned. We
might suppose that instead of falling through the horizon,
particles (and falling observers) encounter an obstruction, a
``brick wall" instead of the horizon. Again the trouble is that an
observer, now the freely falling one, would experience a violation
of a sacred law of nature, this time the equivalence principle.
The horizon according to general relativity is an almost flat
region which should behave like empty space, not like a brick
wall. It seems that there is no way out; at least one or another
observer must witness violations of the usual laws of nature.

At this point one could ask why not just assume neither observer
sees a violation of sacred principles. Just assume that the
infalling information is ``xeroxed" or ``cloned" at the horizon.
One copy would fall through with the infalling observer and the
other copy would get scrambled and radiated with the evaporation
products. Each observer would see the usual laws of nature
respected. The problem is that it potentially violates another law
of quantum mechanics. I call it the ``Quantum Xerox" principle. It
says that quantum information stored in a single system can not be
duplicated. To illustrate the principle consider a single spin
whose $z$ component can take two values. $ \sigma_z = \pm 1. $ We
call the two states $| +\>$ and $|-  \>$. Suppose we had a quantum
xerox machine which could produce a clone of the spin in exactly
the original state. Its action is defined by \bea |+ \> &\to& |++
\> \cr |-\> &\to& |-- \>. \label{zclone} \eea But now consider the
action of the quantum xerox machine on the superposition of states
$|+\> + |- \>$ which represents a spin oriented along the $x$
axis. According to the principles of quantum mechanics the result
must be the  superposition $|++ \> +|-- \>$. This is because
quantum evolution is a linear process. On the other hand the QXM
must clone the original state \be \left(|+ \> +|- \> \right) \to
\left(|+ \> +|- \> \right)\bigotimes\left(|+ \> +|- \> \right)
\label{qm} \ee or \be \left(|++ \> +|-- \> \right) \to |++ \> +
|+- \>  + |-+ \>  + |-- \>. \label{qx} \ee Evidently the quantum
Xerox machine is not consistent with the principle of linear
quantum evolution.

The Quantum Xerox Principle was probably an implicit part of
 Stephen's thinking. I remember that in Aspen sometime around 1992 I
gave a seminar
explaining why I thought loss of coherence was a big problem and
as part of the talk I introduced the  $QX$ principle. A number of
people in the audience didn't believe it at first but Stephen's
reaction was instantaneous: ``So now you agree with me". I said
no, that I didn't but that I was trying to explain to the high
energy people why it was such a serious problem. However, it was
also true that I could't see my way out of the paradox.

\setcounter{equation}{0}
\section{Horizon \C }

{\it{ ``When  you  have  eliminated  all  that  is impossible,
whatever  remains  must  be  the truth,  no  matter  how
improbable  it  is."

Sherlock Holmes}}

The principle of Horizon Complementarity I  interpret to mean that
no observer ever witnesses a violation of the laws of nature. It
is obvious that no observer external to the black hole is in
danger of seeing the forbidden duplication of infalling
information since one copy is behind the horizon. The danger,
pointed out by John Preskill \cite{press}, is that the observer
outside the horizon, call him $B$,  can jump into the horizon
having previously collected the relevant information in the
Hawking radiation. We must now worry whether the original
infalling observer (call her $A$) can send a signal to $B$ so that
$B $ has witnessed information duplication albeit behind the
horizon. The answer is no, it is not possible \cite{gedank}. The
reason is interesting and involves a fact first explained by Don
Page \cite{averageentropy} who realized that to get a single bit
of information out of the evaporation products you must wait until
about half the entropy of the black hole has been evaporated. This
takes a time, in Planck units of order $M^3$ where M is the black
hole mass. It is then easy to see from the black hole metric that
if $B$ waits a time of that order of magnitude before jumping
behind the horizon, then $A$ must send her signal extremely
quickly after passing the Horizon. Otherwise the message  will not
arrive at $B$ before hitting the singularity. Quantitatively the
time that $B$ has after passing the horizon is of order $ \Delta
t\sim \exp{-M^2}$ where all things are given in Planck units.

Now in classical physics an arbitrary amount of information can be
sent in an arbitrarily small time using an arbitrarily small
energy. But in quantum theory if we want to send a single bit we
must use at least one quantum. Obviously that quantum must have an
energy satisfying $\Delta E \sim \exp{M^2}$. In other words $A$
had to be carrying an energy vastly larger than the black hole in
order that $B$ ever witness the forbidden information cloning.
This makes no sense since if A had that much energy it couldn't
possibly fit inside the black hole. We see that a consistent use
of quantum mechanics simultaneously  allows  external observations
to be consistent with  quantum  coherence; the infalling
observations to be consistent with the equivalence principle; and
finally, prevents the observer who chooses to jump in after
accumulating some information from ever seeing information
duplication.

I can now state the principle of Horizon \C . All it says is that
no observer ever sees a violation of the laws of nature. More
specifically it says that to an observer who never crosses the
horizon, the horizon behaves like a conventional complex system
which can absorb thermalize and re-emit all information that falls
on it. No information is ever lost. In essence, the world on the
outside of a horizon is a closed system.

On the other hand a freely falling observer encounters  nothing
out of the ordinary, no large tidal force, high temperature or
brick wall at the horizon.  The paradox of information being at
two places at the same time are apparent and a careful analysis
shows that no real contradictions arise.

 But there is a weirdness to it.
Just to make the point let's imagine a whole galaxy falling into a
huge black hole with a Schwarzschild radius equal to a billion
light years. From the outside, the galaxy and all its unfortunate
inhabitants appear to be heated to Planckian temperature,
thermalized and eventually emitted as evaporation products and all
this takes place at the horizon! On the other hand, the infalling
galactic inhabitants glide through perfectly happily. To them the
trauma only happens at the singularity a billion years later. But
as in the case of certain life-after-death theories, the folks on
the other side can never communicate with us.

It is clear from Horizon \C \ that a revision is needed in the way
we think about information being localized in space-time. In both
classical relativity and in quantum field theory the space--time
location of an event is invariant, that is, independent of the
observer. Nothing in either theory prepares us for the kind of
weirdness I described above.

\setcounter{equation}{0}
\section{The Holographic Principle}

The idea that information is in some sense stored at the boundary
of a system instead of the bulk was called Dimensional Reduction
by 't Hooft and the Holographic Principle by me \cite{hooft,holo}.
The simple argument for the Holographic Principle goes as follows.
Imagine a ball of space $\Gamma$ bounded by a spherical area $A=
4\pi R^2$. Ordinarily we would assume that the entropy in that
region can be arbitrarily large. However if we introduce a cutoff
in space, say at the Planck scale and assume that within each
Planckian volume only a small number of states are possible then
it is natural to assume that the maximum entropy is finite and
proportional to the volume of the region. However it is easy to
see that this can not be so. Imagine an imploding spherical
light-like shell with just enough energy so that the boundary of
$\Gamma$ gets turned into a black hole horizon. The final entropy
of the system is the Bekenstein-Hawking entropy of the black hole
$S_{BH}=A/4G$. By the Second Law the entropy within $\Gamma$ could
not have been larger than this. But by causality this bound can't
depend on  the existence of the infalling light like shell. Thus
it follows that the largest entropy a region of space can have is
given by \be S_{max} =A/4G. \label{smax} \ee A further implication
is that the number of degrees of freedom (binary bits) needed to
describe a region is also proportional to the area. It is as
though the information in the bulk of $\Gamma$ can be mapped to a
set of ``Holographic" degrees of freedom on the boundary. The idea
that the quantum mechanics of a region can be described in terms
of a theory of no more than $A/4G$ degrees of freedom is the
content of the Holographic Principle. It is also very weird but it
is forced on us by  Bekenstein's  observation about black hole
entropy \cite{bek},  the usual interpretation of entropy as
counting states and Holmes' dictum.

\setcounter{equation}{0}
\section{The Ultraviolet/Infrared Connection }

The dominant paradigm of 20th century physics has been that small
distance means high energy or more precisely high momentum. Thus
to probe increasingly small (and presumably increasingly
fundamental) objects, higher and higher energy accelerators have
been built. Again this is a  lesson learned from conventional \qft
. But the more we learn about the combination of gravity and
quantum mechanics the clearer it becomes that this trend will
eventually reverse itself. Imagine trying to probe distance scales
very much smaller than the Planck scale by building the
$Gedankatron$, a collider of colossal proportions, which can
accelerate electrons to momenta way above the Planck scale. Well,
there is no need to build it. We already know what will happen. In
a head--on collision a black hole will form and the black hole
will have a \sc \ radius of order the center of mass energy $M$ in
Planck units. The black hole will evaporate with a temperature of
order $1/M$. Thus the collision will result in the emission of
longer and longer wave length quanta and will probe distance
scales of order $MG$ and not $1/M$. This is the simplest example
of the UV/IR connection. Similar things happen in perturbative
string theory where increasing energy again results in decreasing
spatial resolution.

In order to  clarify the mechanisms behind the UV/IR connection
and show its connection with the \hp \ I will use the famous
example of the AdS/CFT duality of Maldacena \cite{maldacena}.
 Let me quickly remind you of this duality. We begin
with a stack of $N$ D3-branes in ten dimensional string theory.
The branes fill the directions $(1,2,3)$. They are all placed at
the origin in the $(4,5,6,7,8,9)$ directions. We also define
$r^2=(x^4)^2 +...+(x^9)^2$. The geometry of the resulting stack
has a horizon at $r=0$ where the stack is located.

I will not explain in detail the so called decoupling or
near-horizon limit. The literature on the subject is enormous.
Suffice it to say that the D-brane stack is exactly described in
two ways which are dual to one another. The first is by gravity in
an AdS space with horizon at $r=0$. The AdS also has a time-like
causal boundary at $r=\infty$. Although the boundary is an
infinite proper distance from any finite $r$, the geometry is such
that light  takes a finite time to travel to the boundary and
back.

The dual ``Holographic" description is in terms of a 3+1
dimensional conformal gauge theory whose precise details are
unimportant for our purposes. The \qft \ is usually interpreted as
residing of the boundary of the AdS,  infinitely  far from the
horizon. On the other hand the branes that support the open string
field quanta are supposedly  located at $r=0$. There seems to be
some serious confusion about the location of the branes. Are they
at  $r=0$ or $r=\infty$? The answer to this question is closely
related to the paradox of where the information in a black hole
resides.

In the field theory dual the location of the branes in the
$(x^4..x^9)$ space is described by a set of six $N \times N$
matrix valued fields $\phi$. The connection between the location
of a brane and the corresponding field is \be \langle r^2 \rangle
={1\over N^2}\langle Tr \phi^2 \rangle. \label{psit} \ee

Now we can see the problem in localizing the branes. The field
$\phi$ is a canonically normalized scalar field with mass
dimension $1$. As such, it has zero point fluctuations which
render the value of $r^2$ divergent. If taken at face value, this
would say that the branes are located at the boundary at
$r=\infty$.  But this is not the most useful way to think about
the problem. A better way to think about it is to introduce an
ultraviolet  cutoff in the gauge theory. Let the regulator
frequency  be called $\nu$. Then (\ref{psit}) is replaced by \be
\langle r^2 \rangle ={1\over N^2} Tr \langle \phi^2 \rangle
=\nu^2. \label{rpsit} \ee

Once again we see something weird. The location of the branes is
not an invariantly defined quantity. It depends on the frequency
resolution that the observer uses. If the branes are observed in a
way that averages over high frequency oscillations then their
location is near $r=0$. On the other hand if all the ultraviolet
fluctuations are included then the branes are located out near the
causal boundary of AdS and the entire theory in the bulk of AdS is
described by a holographic description in terms of boundary
degrees of freedom. This interplay between short time cutoffs in
the \qft \ and the location of the brane degrees of freedom is
called the UV/IR connection \cite{edlen}.

The UV/IR connection is closely related to the weirdness of \C .
Imagine an observer falling toward a black hole horizon while
blowing a dog whistle. A dog whistle is  just a whistle whose
frequency is beyond the range of human hearing. The freely falling
observer never hears the whistle. But to someone outside the black
hole the frequency is red shifted so that after a while the
whistle becomes audible. In the same way, the ultra-high frequency
fluctuations of an object are invisible under ordinary
circumstances. But as the object approaches the horizon the
external observer becomes sensitive to them. In the example I just
showed you the
 observed location and spread of an object depends on the visibility
of the high frequency fluctuations. Thus the UV/IR connection
provides a mechanism for understanding Horizon \C \ \cite{string}.
Horizon \C \, the \hp \ and the UV/IR connection are different
facets of the same weirdness that characterizes the new paradigm.

\setcounter{equation}{0}
\section{Counting Black Hole Microstates }

Ultimately the question of black hole coherence boils down to
whether or not the formation and evaporation of a black hole can
be described from the outside by an S--matrix. Stephen said no. 't
Hooft and I said yes. Although, for years, many string theorists
sat on the fence about the issue, they really had no choice. The
only mathematically well defined objects in string theory, at
least in a flat background, are S--matrix data. This includes the
complete list of stable objects in the theory and the transition
amplitudes for their scattering. Unstable objects also have
meaning as resonances which can be defined as poles of the
S--matrix in the complex energy plane. If the laws of quantum
mechanics are not violated for a distant observer, black holes are
simply such resonances. It sometimes happens that resonances
become extremely densely spaced. This occurs in nuclear collisions
and also in string theory. In these cases the practical tools are
those of statistical mechanics and thermodynamics.

Complementarity implies that the thermodynamics of a black hole
should arise from a quantum statistical mechanical origin.
Consider the thermodynamic description of a bathtub of hot water.
We specify a few macroscopic variables such as the volume, energy
temperature etc. Of special significance is the entropy. Entropy,
as we know, is a measure of our ignorance of the precise
microscopic details of the tub of water. It measures the logarithm
of the number of quantum states consistent with the macroscopic
description. The existence of an entropy tells us that there is a
hidden set of microscopic degrees of freedom. It doesn't tell us
what those degrees of freedom are but it tells us they are there
and that they can store detailed information  that  our
thermodynamic description is too coarse grained to see. The
principle of Black Hole Complementarity requires  the
thermodynamics of black holes to originate from the coarse
graining of hidden microscopic degrees of freedom and that the
entropy is counting the number of microstates of the black hole.

General relativity does not tell us what those micro-states  are.
But a theory like string theory which is supposed to be a
consistent quantum theory of gravity should tell us and it does
\cite{spec,sen,vafa,curt,gary}. Let us begin with a single string
in very weakly coupled string theory. If the string is excited to
a high state of excitation it typically forms a random walking
tangle. It is natural to describe it statistically.

Let $L$ be the total length of string and $T ={ 1\over {\alpha'}}
$ be the tension. The mass of the string is then \be m=TL.
\label{mass} \ee A good model of the string is to think of it as a
series of links on a lattice. The link size is the string length
$\ls$. Lets suppose that the when we follow the string and it
arrives at a lattice site it can continue in $n$ distinct
directions. We can then count the number of configurations of the
string. The number of links is $L/\ls$ and the number of states is
\be N_{states}=n^{L/\ls}. \label{nstates} \ee Thus the entropy is
\be S={L\over \ls} \log{n} \label{s} \ee or using (\ref{mass})
$S={m \ls} \log{n}.$ The thing to abstract from this formula is
that the entropy is proportional to the mass in string units. The
precise coefficient is determined by the details of string theory.
\be S=c{m \ls}. \label{sm} \ee

Now imagine turning up the string coupling constant. The large
ball of random walking string will begin to experience the effects
of gravity. It will shrink and will gain some negative
gravitational binding energy. Eventually it will turn into a black
hole. This much is obvious. What is less obvious is what happens
if we begin with a neutral black hole and turn off the coupling.
It is clear that it must evolve into a system of free strings in
this limit but how many such strings will be left at the end of
the process? Surprisingly the answer is that the overwhelmingly
most likely final state is a single string! The number of
configurations of one long string vastly outweighs all other
configurations. Thus we can go back and forth from single string
states to black holes.

This suggests the following strategy  \cite{spec}. Start with a
black hole of a given mass $M$. Let us follow it while we
adiabatically turn down the string coupling. The entropy will be
conserved by such an adiabatic process. At the end we will get a
single string of some other mass $m$. If we can follow the process
and determine $m$ we can use (\ref{sm}) to estimate the entropy of
the original black hole.

This strategy has been used to study a large variety of black
holes that occur in string theory \cite{spec,sen,vafa,curt,gary}
including \sc \ black holes in all dimensions, and special
supersymmetric black holes where the precise coefficients can be
obtained. Up to numerical coefficients of order unity, the results
always agree  with the universal formula \be S={area \over 4G}.
\label{bhs} \ee By now string theory has given us great confidence
in the
 counting of black hole microstates and that there is no need to
 look for a new basis for gravitational entropy outside the
 framework of conventional quantum statistical mechanics.

\setcounter{equation}{0}
\section{ \des }

Let me turn now to Stephen's favorite subject, cosmology.
According to the inflationary theory the universe may have
started as  a long lived approximation to \des \ with a very small
radius of curvature, perhaps a couple of orders of magnitude
larger than the Planck radius. Right now observations indicate
that it may end as a \des \ of radius $\sim 10^{60}l_p$. Whether
or not this will ultimately prove to be the case it clearly
behooves us to study and understand the quantum nature of \des .
Once again, Stephen,  together with Gary Gibbons, led the way
\cite{gibhawk}.

\des \ is another example of a space-time that has an event
horizon and therefore an entropy and temperature. Accordingly, we
should expect paradoxical issues of black hole quantum mechanics
to confuse and harass us in this context. The rest of my
discussion will concentrate on the quantum mechanics  of pure \des
. However  before doing so I will remind you of some facts about
\sc \ black holes in AdS \cite{hawkpage}.
 You will see why shortly.
The Penrose diagram for an AdS-\sc \ black hole is a square
bounded on top and bottom by singularities and on the sides by
causal boundaries which are at an infinite proper distance. The
diagonals of the square are the horizons. Note that the boundary
is doubled. As usual, the diagram is divided into 4 regions.
Region I is the exterior of the black hole. The Penrose diagram
has a boost symmetry which in regions I  and III it acts as time
translation invariance but with the convention that in III it
translates in the negative time--like direction. In regions II and
IV it translates in space-like directions.

The AdS black hole has a description in the dual CFT language. It
is simply the state of thermal equilibrium for the CFT above the
temperature corresponding to the Hawking Page transition
\cite{hawkpage}. Since the CFT is a conventional quantum system we
can be sure that the boundary observers see nothing inconsistent
with quantum theory in all possible experiments that they can
perform on the black hole. In particular the region I which is in
causal contact with the boundary and which is described by the CFT
can not experience anything that violates the standard quantum
principles. No information can be lost across the horizon. In this
respect region I is a closed quantum system.

How does black hole complementarity manifest itself in this
system? To see the full implications of exact information
conservation in terms of correlation functions is extremely
complicated. But Maldacena has given a very useful implication
\cite{juan}. Consider the correlation function of two local
boundary operators at widely separated times $t$ and $t'$. For
example the operators could be the gravitational field evaluated
at the boundary and described by the energy-momentum tensor of the
QFT. The correlation function can be computed  by doing a bulk
calculation of the graviton propagator in the AdS-\sc \
background. The answer is that it exponentially tends to zero as
the time separation grows. From the CFT point of view the
exponential decrease has a simple explanation. The thermal
environment leads to dissipation which generally causes
correlations to disappear exponentially. The coefficient in the
exponential is a dissipation coefficient. However, Maldacena
argues that there is something wrong with this conclusion. His
claim is that the correct answer on general grounds is that the
correlation decreases exponentially until it is of order
$\exp{-S}$ where $S$ is the finite black hole entropy. Thereafter
it stops decreasing. If Maldacena is correct it means that
ordinary QFT in the bulk is missing something important and that
something is closely connected to conservation of information
outside the black hole. The argument that the correlation has a
nonvanishing limit was not spelled out in Maldacena's paper but I
will derive it shortly.

Now what does all this have to do with \des ? To answer this we
only need to draw the Penrose diagram for \des . In fact it is
identical to the AdS-\sc \ case! There are very big geometrical
differences but the causal structure is the same. In the de Sitter
case the vertical boundaries of the Penrose diagram are not
spatial infinity but the ``north and south poles" of the spherical
spatial sections. One of these poles we can identify as ``the
observer". The horizontal boundaries are not high-curvature
singularities but are instead the infinitely inflated past $I^-$
and the infinitely inflated future $I^+$.

The metric for \des \ is \be ds^2 =R^2 \left( dt^2 -\cosh^2(t)
d\Omega_d^2 \right) \label{global} \ee where $R$ is the radius of
the \des \ and $d$ is the dimension of space.

If \des \ exists in a quantum theory of gravity, perhaps the most
urgent question is what are the mathematical objects that the
theory defines. This is particularly important for string theory.
In flat space the answer is S-matrix data. In AdS it is boundary
correlators of the gravitational and other bulk fields. In both
cases we go to the boundary of the world and define vertex
operators whose correlators are the ``definables". The obvious
suggestion for \des \ is that we do the same thing. The boundary
of \des \ is the union of $I^-$ and $I^+$. This suggests that the
definables consist of S--matrix--like elements which relate
initial states on $I^-$ to final states on $I^+$
\cite{dscft,witten,volo}. I have used the term definables
\footnote{Witten has used the terms computables and
meta-observable for the same objects.} and not observables for the
reason that no one can ever observe them. The argument is closely
connected to that  of Section 3 about why information duplication
can't be detected behind a black hole horizon. In that case there
was just not enough time for any observer to collect information
from the observers $A$ and $B$ before reaching the singularity.
The similarity of $I^+$ to the singularity of a black hole means
that two observers at fixed spatial  coordinates can not transmit
the results of measurements to one another once global time gets
too late.  Accordingly correlators of fields at the future
boundary are unobservable.

In the black hole case it is not just that quantum correlators on
the singularity are un-measurable. Just their mere $mathematical$
existence would require us to trace over them in defining final
states. It is of obvious importance to know if the same is true in
\des . In other words can similar arguments show that the
existence of \des \ boundary correlators will lead some observer
to see a violation of the laws of nature? I will argue that the
answer is yes. But first we need to formulate an appropriate \C \
principle.

In the black hole case, a convenient starting point for discussing
observations outside the hole is the presentation of the geometry
in \sc \ coordinates. The distinguishing features of these
coordinates are that they are static and that they only cover the
region  on the observer's side of the horizon. In \des \ an
observer means a time-like trajectory that begins on $I^-$ and
ends on $I^+$. All such pairs of points are related by symmetry
and it is always possible to choose them at the same spatial
location. Thus we can always choose the observer to be the $r=0$
edge of the Penrose diagram. By analogy with the AdS/\sc \ case
the diagonals of the square Penrose diagram are horizons and the
region on the observers side of the horizon is a triangle. On this
region it is possible to choose static coordinates so that the
metric takes the form \be ds^2= R^2 \left( (1-r^2)dt^2
-(1-r^2)^{-1}dr^2 -r^2 d\Omega_{d-1}^2 \right). \label{stat} \ee
This geometry has a horizon at $r=1$ which now surrounds the
observer. As usual, the horizon has an entropy \be S=area/4G =\pi
R^2 /G \label{sdes} \ee and a temperature. The proper temperature
at a point $r$ is given by \be T(r)={1\over 2 \pi R \sqrt{1-r^2}}
\label{tofr} \ee As for Rindler or \sc \ space,  the temperature
diverges near the horizon. The observer at $r=0$ experiences a
more modest temperature.

The statement of Horizon \C \ is fairly obvious. An observer at
$r=0$ sees the world as a closed finite system at a non--zero
temperature. Once again, closed means that no information is lost
and the system obeys the rules of quantum mechanics for such
closed systems \cite{tom}. What are the implications of the
complementarity principle and especially the conservation of
information? As in the similar AdS/\sc \ case there are
implications for the long time behavior of correlators. However
before discussing them I want to consider the behavior of field
correlators in ordinary \qft \ in the fixed \des \ background. For
this purpose I return to global coordinates.

For simplicity I will use the example of a massive scalar field
$\Phi$ with mass $\mu/R $. The field equation is easily worked out
and the asymptotic behavior of $\Phi$ is seen to be \cite{dscft}
\be \Phi \to \exp{-\gamma |t|} \label{phi} \ee as $t \to \pm
\infty$. The ``diffusion constant" $\gamma$ is given by \be
\gamma=d \pm i\sqrt{\mu^2- {d^2\over4}}. \label{gam} \ee The
complex value of $\gamma$ has a simple meaning in terms of damped
oscillations. Depending of the value of $\mu$ the oscillations of
the field are either under-damped or over-damped. The real part of
$\gamma$ is always positive.

Now let us return to static coordinates and consider the
correlator at $r=0$ and large time separation. Time translation
invariance in the static patch together with (\ref{phi}) requires
the correlator to behave like \be F(t)=\langle \phi(t)\phi(t')
\rangle \sim \exp{- \gamma|t-t'|} \label{foft}. \ee Thus,  as in
the AdS/\sc \ black hole the correlator evaluated by naive \qft \
tends to zero exponentially with time. As we will see, this is
inconsistent with the finite entropy of the static patch.

\setcounter{equation}{0}
\section{ \ Correlations in Finite Entropy Systems }

Let us consider an arbitrary system described by Hamiltonian $H$
at temperature $1/\beta$. I will not assume anything about the
total number of states of the system but only that the thermal
entropy is finite, $S=finite$. An immediate implication is that
the spectrum is discrete and that the level spacing is of order
\be \Delta E \sim \exp{-S}. \label{level} \ee Let's label the
energy levels $E_m$.

Now  consider a correlator of the form \cite{lisa} \be \langle
\f(0) \f(t) \rangle \equiv F(t). \label{ff} \ee It is defined by
\bea F(t)&={1 \over Z}&Tr e^{-\beta H} \f(0) \f(t) \cr
    &=& \sum_{mn} e^{-\b E_n} |\f_{nm}|^2 e^{it(E_m-E_n)} .
    \label{trff}
\eea For simplicity I will assume that the diagonal matrix
elements of $\f$ vanish. This implies that the time average of
$\f$ also vanishes.

I want to determine if $F$ tends to zero as $t \to \infty$. A
simple way to do this is to compute the long--time average of
$F^\ast F$. \be L= \lim_{T\to \infty} {1\over 2 T} \int_{-T}^T
F^\ast(t)F(t). \label{lta} \ee If $F \to 0$ then $L=0$.

Using (\ref{trff}) it is straightforward to compute L; \be L={1
\over Z^2}\sum_{mnrs} e^{-\b(E_m+E_r)} |\f_{mn}|^2 |\f_{rs}|^2.
\label{Lequals} \ee This is a positive definite quantity. We have
therefore proved that the correlation function does not go to
zero.

It is not hard to estimate $L$ \cite{mark}. We use the fact that
the level spacing is of order $\exp{-S}$ which implies that the
matrix elements of $\f$ are of the same order of magnitude. The
result is \be L \sim \exp{-2S} \label{emts} \ee indicating that at
asymptotic times the average magnitude of the correlator is of
order \be |F| \sim \exp{-S} \ee which agrees with Maldacena's
guess.

However, this does not mean that the correlator tends smoothly to
an exponentially small constant. In fact it does something quite
different. It becomes ``noisy". Typically the fluctuations are
small, of order $\exp{-S}$. But if you wait long enough large
fluctuations occur. On sufficiently long time scales the
correlation function return to close to its original value. In
 \cite{lisa}  the reader can find plots of the results of some numerical
studies
done by Lisa Dyson and James Lindesay  \cite{lisa} which show the
typical behavior. The large scale fluctuations are the quantum
version of Poincare recurrences in classical mechanics. They are
just an indication that in a closed finite system, if you wait
long enough everything will eventually happen and not just once
but in eternal repetition.

What is the significance of this fact and why is it missed by \qft
\ in the \des \ background? First of all it means that any
formulation of quantum gravity in \des \ that relies on the
existence of well defined asymptotic fields must fail. Among these
I include the possibility that string theory might provide a set
of scattering amplitudes relating asymptotic states on $I^{\pm}$.
Another version of \des \ quantum gravity is the so called dS/CFT
correspondence inspired by the AdS/CFT duality.  This theory
attempts to organize boundary correlators into an Euclidean \qft .

A question which often comes up in any discussion of \des \ is how
the static patch observer should think about objects
(conventionally called elephants)  which enter and and leave the
static patch through the diagonal past and future horizons. Such
events are described by very rare fluctuations which materialize
among the large number of low frequency degrees of freedom very
near the horizon. In other words they are the very intermittent
large fluctuations that can be see in the numerical simulations of
\cite{lisa}. Such elephants can not be tossed in from $I^-$. To
understand why, let's think about the elephant in its own frame.
Because of the thermal fluctuations  in \des \ an elephant will
eventually be thermalized and turned into black body radiation.
Since $I^-$ is an infinite proper time in the past any elephants
that were present in the initial conditions have long been
thermalized by any finite time. Elephants in \des \ are thermal
fluctuations which form by chance and then evaporate.

Finally, why does \qft \ in the  \des \ background miss these
fluctuations? The answer is simply that in \qft \ the entropy of a
horizon has an ultraviolet divergence due to modes arbitrarily
close to the horizon. These modes provide an infinite number of
low frequency oscillators which effectively make the system open.
But we know that most of these modes must be eliminated in a
proper theory of quantum gravity since the entropy of horizons is
finite.

If boundary correlators and S-matrix elements can not exist in
\des \, what then are we to base a quantum theory on? Here things
are much more obscure than in  AdS. A possible answer is to define
a theory in the closed static patch. But, as  emphasized by Bohr,
the clearest use of quantum mechanics is when we can separate the
world into a system and a large or classical apparatus (observer).
That is just what we can not do in a closed finite thermal system
like \des . In fact the whole discussion of static patches being
defined by infinitely long lived observers in \des \ is internally
inconsistent. Like elephants, any real observer is subject to
bombardment by the thermal radiation and will become degraded.
Exactly how to think about quantum mechanics is far from clear.
Perhaps eternal \des \ is just a bad idea and will go away.
Perhaps not.

So you can see that Stephen and I have had much to disagree about.
These disagreements have been incredibly exciting and certainly
provided the high point of my own career. There is no doubt that
Stephen's remarkable question about black holes has left an
indelible mark on me  but more importantly, on the future of
physics.

Happy birthday, Stephen and best wishes for many more.

\end{document}